\documentclass{elsart}
\usepackage{graphicx,epsfig,amssymb}
\journal{Physica D}

\begin{document}

\begin{frontmatter}
\title{Prisoner's dilemma on co-evolving networks under perfect rationality
}
\author[COSY,ATI]{Christoly Biely},
\author[COSY]{Klaus Dragosits},  
\author[COSY,ATI,cor1]{Stefan Thurner}
\corauth[cor1]{
Corresponding author, present adress: 
Complex Systems Research Group, HNO, Medical University of Vienna,
W\"ahringer G\"urtel 18-20, A-1090 Vienna, Austria;
email: thurner@univie.ac.at
}
\address[COSY]{
Complex Systems Research Group\\
HNO, Medical University of  Vienna\\
W\"ahringer G\"urtel 18-20, A-1090 Vienna, Austria
}
\address[ATI]{
Atominstitut der \"Osterreichischen Universit\"aten\\
Stadionallee 2, A-1020 Vienna, Austria
}

\title{}

\author{}

\address{}

\vspace{-3.0cm}
\begin{abstract}
We consider the prisoner's dilemma being played repeatedly on a
dynamic network, where agents may choose their actions as well as their
co-players. This leads to co-evolution of network structure  
and strategy patterns of the players. Individual decisions are made {\it fully rationally} and are based  
on  {\it local information} only. They are made such that links to defecting agents are
resolved and that cooperating agents build up new links. The exact  
form of the updating scheme is motivated by profit maximization and not by  
imitation. If players update their decisions in a synchronized way the system  
exhibits oscillatory dynamics: Periods of growing cooperation (and total linkage) alternate with  
periods of increasing defection. The cyclical behavior is reduced and the system  
stabilizes at significant total cooperation levels when players are less synchronized.
In this regime we find emergent network structures resembling  
'complex' and  hierarchical topology. The exponent of the power-law degree distribution ($\gamma\sim8.6$)  
perfectly matches empirical results of human communication networks.
\end{abstract}


\end{frontmatter}

\section{Introduction}

The recent years have seen a drastic increase of interest in  
understanding the emergence of complex structures in nature and society. In this  
context, network theory has played an important role because it provides a topological  
substrate of discrete real-world interactions \cite{barabasi,doro}.
The science of networks  basically follows two main lines of  
research: On the one hand the formation of network structures is studied involving predominatly {\it  
topological} parameters (e.g. preferential attachement \cite{barabasi}).
On the other hand, dynamical processes \emph{on} networks with fixed topology have
been studied, see e.g. \cite{bosa}.
In this context, also game theoretic models -- in particular the prosners'
dilemma -- have been analyzed on distinct network topologies.
So far not much work has focused on the co-evolution of topological  
\emph{and} internal degrees of freedom, see however \cite{ito_kaneko} for a quite  
general starting point. In the article we want to specifically address this matter by  
discussing a model where internal and topological degrees of freedom mutually influence each  
other. The model is based on the prisoner's dilemma (PD) \cite{axelrod84} -- one of  
the most impressive ways of illustrating situations of human interactions where mutual  
trust is beneficial, but egotism leads to a breach of promise.
The fundamental interest in the PD arises from its applicability in a  
variety of fields, ranging from physics and  biology to economics and finance \cite 
{bokros,huberman,miller}. It  is of particular interest in constitutional economics \cite 
{brennan85,buchanan75,tullock}.

The central point in the PD dilemma is the payoff matrix, whose
specific form reads for the payoff of one of two players, say player  
$i$,
\begin{equation}
\label{payoff_matrix}
P_{ij}=\left(
\begin{array}{cc}
R & S\\
T & P  \end{array}
\right)
\qquad
P_{ij}=\left(
\begin{array}{cc}
3 & -1\\
4 & 0  \end{array}
\right)  \quad.
\end{equation}
Each player has two options: she can  defect (D) or cooperate (C).
Mutual cooperation yields the highest total payoff -- giving each of  
the players an equal payoff of $R$ (reward); this is the  optimal strategy when seen from a 'global' point of view.
If one of the players defects while the other cooperates, the highest  
attainable individual payoff -- i.e. the temptation, $T$ -- goes to the defector  
and the cooperator receives the lowest possible payoff, the sucker's payoff, $S$. The  
cooperator would have been better off if he would have defected, thus receiving the payoff $I$.  
In a one-shot game, the dilemma holds as long as the entries of the payoff  
matrix, Eq. (\ref{payoff_matrix}), satisfy $S<I<R<T$ and $2R>T+S$.

Much research has concentrated on spatial aspects of this game, 
initially introduced in the pioneering work of Nowak \& May \cite{NowakandMay}.  
In their work, players are located on a square lattice and play repeatedly (!)
with their neigbours. Cooperation is made possible by the assumption, that every agent imitates 
his neighbours in such a way, that they \emph{synchronously} choose the
actions of the neighbours who got the highest payoff in 
the last turn. This specific rule of evolution leads to non-trivial complex spatio-temporal
dynamics. It has been noted that the model is seriously troubled by the fact
that an asynchronous update
of strategies leads to the break-down of  cooperation \cite{huberman_new}.

Carrying the discussion to more complex structures, the intial model of Nowak \& May and
slight adaptions thereof have been extensively studied under the aspect of 
different interaction topologies, see \cite{review} for a
review of recent developments.
In \cite{santos} it was shown, that for a variety of dilemmas (including the
PD), heterogenous networks (e.g. scale-free networks) favor the emergence of
cooperation. 
The role of hierarchical lattices was elaborately discussed in \cite{szabo2}. Interesting aspects
of the PD on random graphs have been analyzed in \cite{duran_mulet}; the role of 
small world networks was tackled in \cite{Kuperman}. 
Effects of entries in the payoff matrix and addition of noise have been
examined on different types of two-dimensional lattices \cite{szabo3}. 
Other topology related topics, such as the role of an
'influential node' \cite{Trusina} and optional participation \cite{szabo1} have been examined as well. 
In essence, a vast number of possible topologies and formulations 
has been studied. However the networks are static and do not
dynamically evolve. 

It is important to note that 
internal sanctions (refusal/termination of links) and positive
feedback mechanisms ('preferential' choice of cooperating agents)
are both directly related to variability of the underlying network and 
may play an important role in real-life situations.
Few, but promising works have brought forward research towards this end in the
recent years: In \cite{ashley_smucker}, 
players keep a running average of payoffs obtained from each other
player in a simulated tournament. These averages effectively determine whom to approach and whom to accept
as co-player in the future. The strategies are basically determined by the 'genetic code'
of the players and altered by crossover and mutation during the tournament.
In \cite{Hauk} it was examined how preferential partner selection influences the
performance  of fixed strategies, thus serving as a starting-point 
for models in which players can also choose their strategy.
Recently, results where the evolution of strategies is driven by imitation and coupled with evolution of the interaction
network have been presented \cite{zimmermann04} (see
\cite{eguiluz_social} for a discussion of sociological aspects). Keeping the total
number of links fixed, the authors found that the system may reach a
steady state where agents predominantly cooperate. 

In the present paper, we want to study the  
formulation of the prisoner's dilemma including both: network
dynamics and choice of actions.
Close to the 'original formulation' of the PD rationality (not imitation) will be the basis of individual 
decisions. We  also keep information-horizons local,  
thus basing our model-dynamics on a quite strict interpretation of
the \emph{homo oeconomicus}.
We show that 
the dilemma of overall defection can be overcome despite rationality, thus
resolving the prisoner's dilemma:
Even within a population of selfish agents (who maximize their  
expected payoffs
for the next turn) cooperation emerges without the necessity of external
rules, imitative behaviour or the introduction of strategies.
As no memory of the agents is involved, our model
also remains also temporarily local  and incorporates co-evolutionary
dynamics which display interesting collective phenomena and nonlinear  
dynamics.
It is especially intriguing that the resulting cooperation networks show
the same power-law exponents as those found in real communication  
networks,
in particular in mobilephone-call networks \cite{mobile}.

The paper is structured as follows:
The model is presented
in Section 2. In Section 3 results based on a numerical  
implementation of the model are
presented. The influence of  model parameters are discussed as is the  
structure of the networks obtained.
Finally, a discussion of the main results is
provided in Section 4.

\section{The model}

We consider a network with a fixed number of $N$ agents/players with a variable number
$L$ of links between them, where linked agents play the PD-game. 
By $\tilde{N}_i(t)$ we denote the set of  $l_i(t)$ neighbours of agent $i$ on the network at
timestep $t$.
The actions of the agents  are encoded in two-dimensional unit-vectors, i.e.
$a_{i}(t)=a^c=(1,0)$, if agent $i$ cooperates and 
$a_{i}(t)=a^d=(0,1)$, if agent $i$ defects. 
We assume that agent $i$ has full knowledge about the chosen strategy $a_j(t)$
and the payoff $P_j(t)$ of each of her
neighbours $j$, but no knowledge about these quantities for all the other
players (local information).
At each timestep,
agent $i$ performs an update of his action and local neighbourhood with
probability $p_u$. Thus, decisions for chosing neighbours and the actions are
made simultaneously by an average
 number of $N^{update}\approx{}p_u{}N_{tot}$ agents. 
For $p_u=1$, the decisions of the agents are fully synchronized, whereas 
$p_u<1$ automatically includes the important case of
asynchronous updates \cite{huberman_new}.
Once chosen for update, agent $i$ performs maximization of her expected
payoff in the next round, i.e. she maximizes
\begin{equation}
\label{profmax}
\bar{P}_i(t+1)=\sum_{j\epsilon\tilde{N}_i(t+1)}{{a}_i(t+1)P_{ij}\bar{a}_j(t+1)}\quad.
\end{equation}
Here, $P_{ij}$ denotes the payoff matrix, Eq. (\ref{payoff_matrix}), and $\bar{a}_j(t+1)$ the expected
action of neighbour $j$. The
preceding action is taken as reasonable expectation value\footnote{This step may be critized as being inductive. However, for $p_u<1$
player $i$ will know that his neighbours will keep their strategy on
average. Thus our argument is inductive only for $p_u=1$.}, $\bar{a}_j(t+1)=a_j(t)$. 
In Eq. (\ref{profmax}), profit-maximization of agent $i$ is performed by adjusting the future action
$a_i(t+1)$ as well as the future (expected) neighbourhood $\tilde{N}_i(t+1)$.
At this point, further substantiation of the network-dynamics is inevitable.
In the following we will specifiy detailed rules 
concerning the individual updates $\tilde{N}(t)\rightarrow{}\tilde{N}(t+1)$.

First, we assume that agents cancel a link if the payoff with the 
respective co-player is smaller than, or equal to zero, i.e. if the link
does not pay off
. A unilateral decision for link-cancellation
will suffice to break off of a relationship.
The maximum number of links agents may cancel in one period is limited by a model-parameter $\alpha$;
the neighbourhood after cancellation of $\alpha$ links is depicted 
by $\tilde{N}_i(t+1;\alpha)$. 
The parameter $\alpha$ models 
the maximum number of relations (to defectors) one
is willing to quit per timestep, thus describing the 'sanction-potential' in
the system. 

As far as the creation of new links is concerned, 
we conceive that only agents who have chosen to cooperate have the possibility to establish new
links. 
We make this assumption for 2 realistic reasons:
On the one hand, one could assume that the players enter commitments about their future strategies
(a typical element of cooperative game theory). Then, their strategies 
will practically be known in advance by potential co-players 
and it is reasonable that links offered by agents who anounce to defect will not be
accepted.
On the other hand, it is tempting to conjecture that a mechanism of 'recommendation'
governs decisions of acceptance or refusal of new link-proposals:
It is only rational (and we have assumed rationality of the players)
that next-to-nearest neighbours of $i$ will 'poll' neighbours in
common with $i$ to get an idea about $i$'s strategy
and that they would only accept a link offered by $i$ in case $i$ is 'known' to
cooperate. 
In this framework it is natural that unilateral decision to establish a new
connection will suffice, i.e. if player $i$ cooperates and decides to link
with player $j$, the link will be accepted with certainty as $j$ has no reason
to refuse (the new link will allow him to pocket in  a riskless profit). 
Together with the unilateralism in cancellation of links this makes
complicated 'matching' of the agents' decisions unnecessary.
We limit the maximum number of links which may be established per timestep
by the parameter $\beta$; thus we incorporate 
some constraint on resources which can be spent to establish new
links\footnote{In future work, it could be interesting to study the effect of  costs by
 making $\beta$ proportional to some measure of payoff.}. 
The new neighborhood associated to establishment of $\beta$ links is depicted by
$\tilde{N}_i(t+1;\beta)$.
In summary, the parameters $\alpha$ and $\beta$ can also be seen as 'agility' or willingness to
change partners upon new information.

With the given specification of network-dynamics, we can formulate the
maximization of the payoff, Eq. (\ref{profmax}), in the following way: Each of the $N^{update}$
agents calculates the expected payoff in case of cooperation, 
\begin{equation}
\label{est_coop}
\bar{P}_i^c(t+1)=\sum_{j\epsilon\tilde{N}_i^c(t+1)}
a_i^{c}
P_{ij}{a}_j(t)=\sum_{j\epsilon\tilde{N}_i(t+1;\alpha)}
a_i^{c}P_{ij}{a}_j(t)
+\bar{P}_i^{add}(t+1;\beta)
\end{equation}
and the expected payoff in case of defection, 
\begin{equation}
\label{est_defe}
\bar{P}_i^d(t+1)=\sum_{j\epsilon\tilde{N}_i^d(t+1)}
a_i^{d}
P_{ij}{a}_j(t)=\sum_{j\epsilon\tilde{N}_i(t+1;\alpha)}
a_i^{d}P_{ij}{a}_j(t)
\end{equation}
and will choose the strategy with the higher payoff (if payoffs are equal
the strategy is chosen at random).
$\tilde{N}_i^c(t+1)$ and $\tilde{N}_i^d(t+1)$ denote the 'expected'
neighbourhoods for the two cases. For cooperation they can be written  
as $\tilde{N}_i^c(t+1)=\tilde{N}_i(t+1;\alpha)\cup\tilde{N}_i(t+1;\beta)$
and for defection as
$\tilde{N}_i^d(t)=\tilde{N}_i(t+1;\alpha)$
($A\cup{}B$ denotes the 
union of sets $A$ and $B$). 

Now, the missing piece for determining the action in the next timestep 
is the estimation of the additional expected payoff $\bar{P}^{add}_i(t+1;\beta)$, which can be
acquired due to new links. 
To do so, each agent performs an evaluation of the neighborhood only using information
about nearest neighbors, as illustrated in
 Figure \ref{pic1}: Agent $i$ first evaluates her payoff
obtained from the set of neighbours he and $j$ have in
common, denoted by $P_{\tilde{N}_{(i,j)}}$.
He can then subtract this payoff from $j$'s total payoff, $P_j(t)$, to obtain
an approximation of the profit $j$ gains from the neighbours they do
\emph{not} have in common, denoted by $N_{(i,j)}^{non}$. Weighting
this estimate with the fraction $\beta/N_{(i,j)}^{non}$ and averaging over all neighbours $\tilde{N}_i(t)$, 
agent $i$ obtains the expected additional payoff he receives when 
establishing $\beta$ random links to next-to-nearest
neighbours
\begin{equation}
\label{est_add_payoff}
\bar{P}_{i}^{add}(t+1;\beta)=\frac{1}{\tilde{l}_i(t)}\sum_{j\epsilon\tilde{N}_i(t)}\Theta{}(P_j(t))\left(P_j(t)-P_{\tilde{N}_{(i,j)}}(t)\right)\frac{\beta}{N_{(i,j)}^{non}}\quad,
\end{equation}
which completes the model.
Regarding the sum in Eq. (\ref{est_add_payoff}), we found 
it realistic to confine the summation over $\tilde{N}_i(t)$ to
a summation over a subset of $\tilde{N}_i(t)$, namely to the first-nearest
neighbours of $i$ who have a payoff $P_j(t)>0$: It would be barely rational
for an agent to build up links for which he knows that the expected
payoff is negative on average. This is also why $\tilde{l}_i(t)$ denotes the 
effective number of neighbours contributing in the sum.
Although the numerical results given in the
next section refer to this specific formulation of the model, dropping the
$\Theta(P_j(t))$-term practically gives the same results (we will discuss the
miniscule effect of this term below).
We also note, that Eq. (\ref{est_add_payoff}) gives the highest possible value only if all
next-to-nearest neighbours cooperate\footnote{It is easy to see that, if $j$ does
not cooperate, $i$ can adjust for this via calculating his payoff on $\tilde{N}_{(i,j)}$ assuming
defection and correcting $P_j(t)-P_{\tilde{N}_{(i,j)}}$ for the 
difference in the payoff matrix between defection and cooperation.}.

\begin{center}
{\renewcommand{\baselinestretch}{1.0}\normalsize 
\begin{figure}
\begin{center}
\resizebox{0.35\textwidth}{!}{\includegraphics{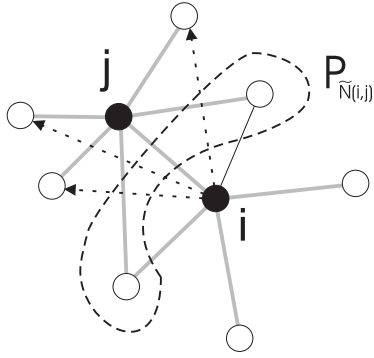}}
\end{center}
\caption{\label{pic1} Illustration for the notation of variables characterizing the
neighborhood of players $i$ and $j$. The players have two neighbors in
common; the corresponding set is denoted by $\tilde{N}_{(i,j)}$. Agent $j$ has
$N_{(i,j)}^{non}$ neighbors not in common with $i$ which are potential new
coplayers for $i$.
The payoff player $i$ obtains from the set of neighbours $\tilde{N}_{(i,j)}$ is
denoted $P_{\tilde{N}_{(i,j)}}$.}
\end{figure}
}
\end{center}

After the evaluation of Eqs. (\ref{est_coop}) and (\ref{est_defe}) has taken place, 
the strategies of the $N^{update}$ agents are updated at the end of each timestep and links are
removed and built up. 
We have already discussed that there is no need for a complicated
'matching'-procedure, as dynamics are governed by unilateral decisions (of
course, it will also happen that two players both decide to play with
each other in the next turn). 
Finally we note, that an agent is randomly wired with one link into the
network if he happens to loose all his links during time evolution of the system.

\section{Results and Discussion}
\label{secresults}

In the following we discuss the model in dependence of the three main parameters -
$p_u$, $\alpha$ and $\beta$. 
As a starting point for our simulations,
 we generated random networks \cite{erdoes_original} of size of $N=10^3$.
Our
simulations have clearly shown that the dynamics and the emerging interaction
networks do not depend on the initial configurations: The system converges relatively fast
towards its attractors (repulsors). 
Simulations have been typically performed for $10^5$ timesteps,
providing accurate statistics.
If not stated otherwise, the payoff matrix was chosen in the specific form given in
Eq. (\ref{payoff_matrix}).
We also studied the effect of changing
the entries in $P_{ij}$ which will be discussed below.

\begin{center}
{\renewcommand{\baselinestretch}{1.0}\normalsize 
\begin{figure}
\begin{center}
\resizebox{1.0\textwidth}{!}{\includegraphics{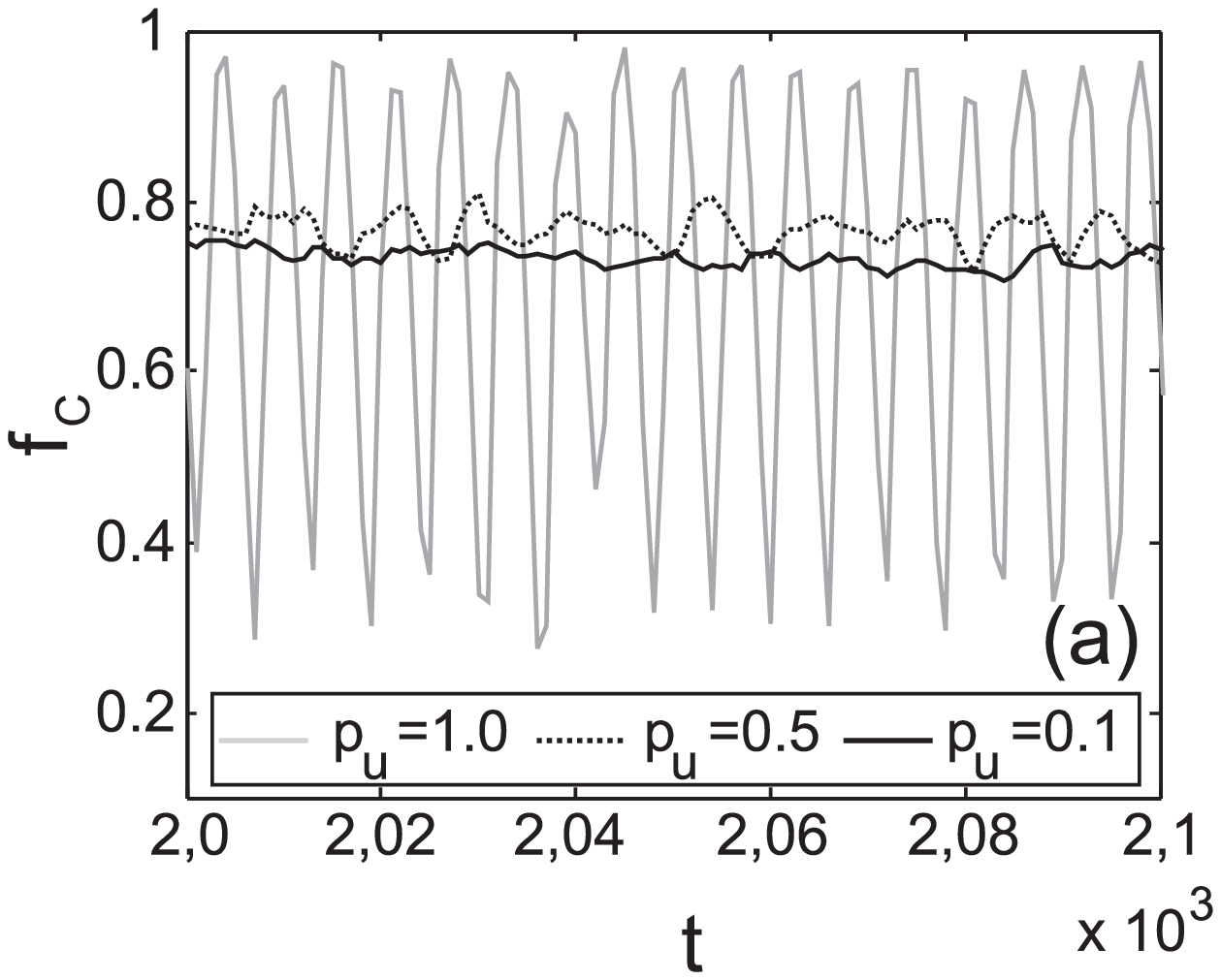}\hspace{1.0cm}\includegraphics{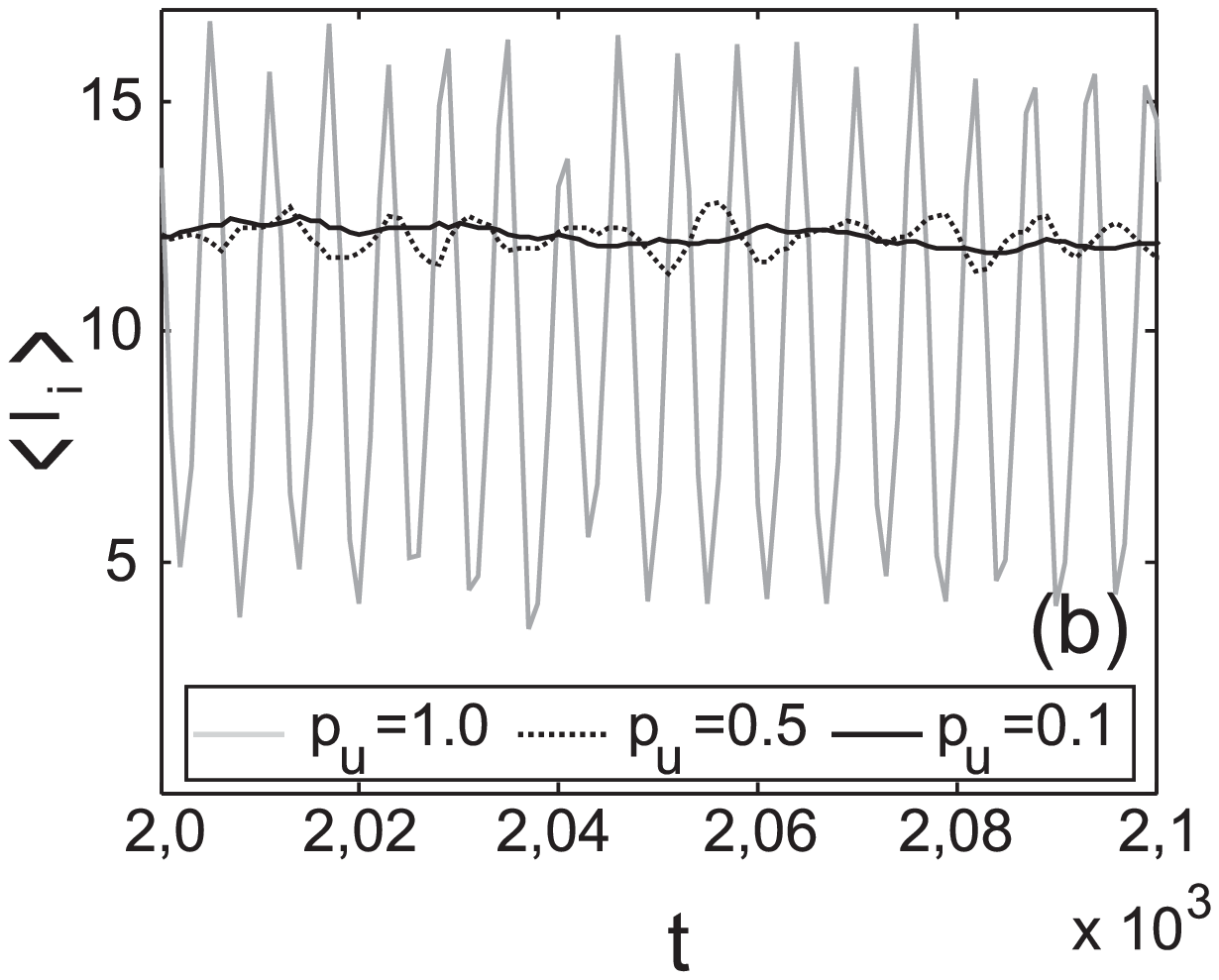}
\includegraphics{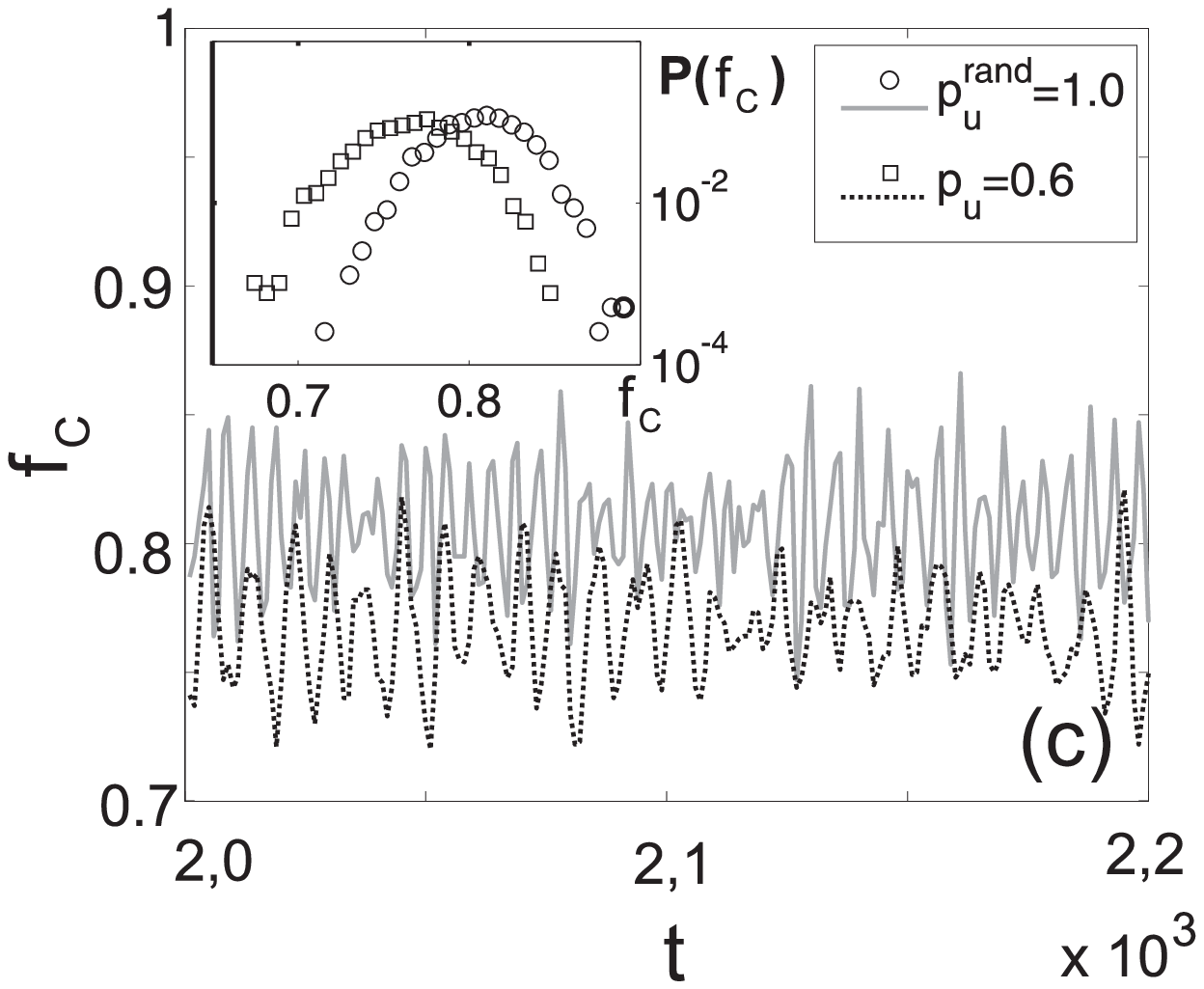}}
\end{center}
\caption{\label{pic2} (a) Time-series of the fraction of cooperating agents $f_c$ for different values
of the update-probability $p_u$ ($p_u=1.0$, $p_u=0.5$, $p_u=0.1$), showing decreasing regularity.
(b) Time-series of the average number of links $\langle{}l_i\rangle$ per agent
for the same values of the update-probability $p_u$. Clearly, the time average $\langle{}l_i\rangle$ for $p_u=0.1$ and $p_u=0.5$ is
considerably above the case for $p_u=1$, indicating the stabilisation of the
corresponding network. (c) Comparing the time-series pertaining to $p_u=0.6$
of the model to $p_u^{rand}=1.0$  of a 'random' formulation of the model
described in the text. The inset shows the empirical distribution of both time-series.
}
\vspace{1.0cm}
\end{figure}
}
\end{center}

\subsection{Properties of cooperation time-series}
To discuss basic properties of the time-series, Figure \ref{pic2} depicts the fraction of
cooperating agents (denoted by $f_c(t)\equiv{}N_c(t)/N$) and the average linkage $\langle{}l_i(t) \rangle{}=L(t)/N$
of a particular simulation for $\alpha=\beta=6$ and various values of $p_u$. 
For $p_u=1$, oscillations with a comparably high
amplitude are observed. Also the average linkage  oscillates strongly between a minimum of about 4 and a maximum of
13 links per agent. The reason for the cyclical behavior of the system can be easily understood: 
In the states corresponding to low $f_c$, linkage has been reduced to an extent motivating the agents to build up links
again. In  configurations with high $f_c$, the majority of agents has collectively acquired a state of maximum
linkage, meaning there is no more motivation to cooperate in our rational
setting. 
This can be easily understood since agent $i$ only cooperates
as long as the condition 
\begin{equation}
\label{eqmax}
\bar{P}_i^{add}(t+1)+l^c_i(t)R+l^d_i(t)S>l^c_i(t)T+l^d_i(t)I
\end{equation}
is fulfilled, i.e. as long as the payoff expected from cooperation is larger
than the payoff expected for defection. Here, $l_i(t)=l_i^c(t)+l_i^d(t)$,
where $l^c_i(t)$ denotes the number of links to
cooperating neighbours and $l^d_i(t)$ the number of defecting neighbours. 
If all neighbours of agent $i$ cooperate ($l_i^d(t)=0$), 
he will defect as soon as 
$l_i(t)>\bar{P}_i^{add}/(T-R)$
where
$\bar{P}_i^{add}$ is bound from above by $\beta{}R$. Therefore, for the parameters
chosen here agents in a cooperative
neighborhood will only cooperate as long as $l_i(t)<18$.
The observed rapidity of the oscillations becomes clear, when one considers that the
agents may build up $\beta=6 $ links per move and therefore reach 
$l_i^{max}$ comparatively fast. 
By lowering $\alpha$ and $\beta$ the amplitudes reduce, as intuitively expected (not shown).
\begin{center}
{\renewcommand{\baselinestretch}{1.0}\normalsize 
\begin{figure}
\begin{center}
\resizebox{0.5\textwidth}{!}{\includegraphics{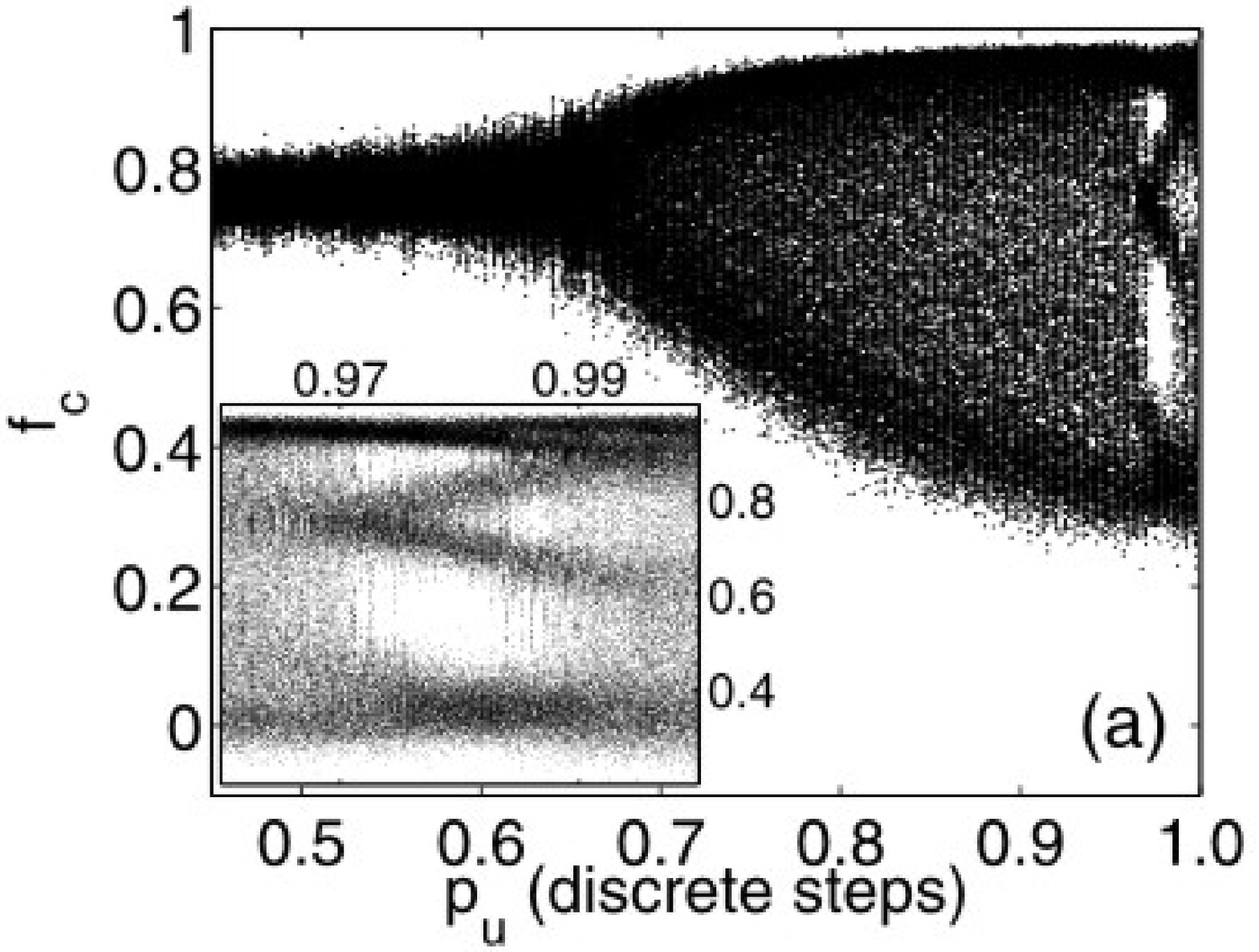}}
\resizebox{0.45\textwidth}{!}{\vspace{1.0cm}\includegraphics{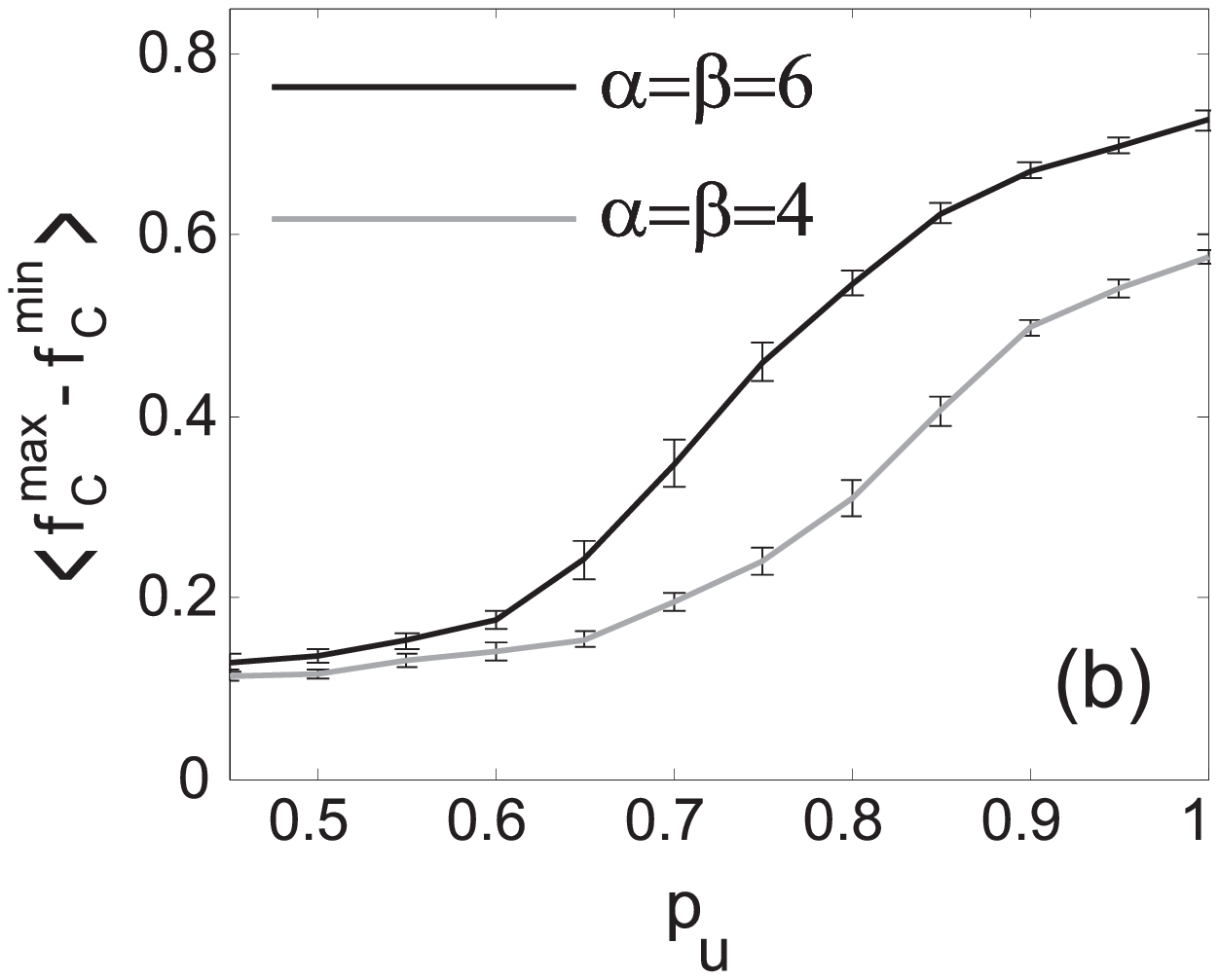}}
\end{center}
\caption{\label{bifurc} 
(a) States $f_c$ visited by the dynamics as a function of the update probability
  $p_u$. $p_u$ is changed in discrete steps. For each $p_u$,
  500 consecutive states $f_c$ are plotted. (b) Range
  $\langle{}f_c^{max}-f_c^{min}\rangle$ of the oscillations as a function of
  $p_u$, averaged over $500$ independent 
  realizations of time-series.
}
\vspace{1.0cm}
\end{figure}
}
\end{center}

\subsection{Dependence on update-probability $p_u$}

In reality, agents are not 
infinitely fast in assessing new information in their
surrounding, as they need time to adopt and employ decisions. 
It is  known from previous studies \cite{huberman_new} that 
asynchronous update can strongly influence observed dynamics and level of cooperation.
Considering this by lowering $p_u$, the oscillations in the overall population
are increasinlgy damped, indicating that the network is stabilized in
comparison to the $p_u=1$ case (see Figure \ref{pic2}). In contrast to the rapid update mode
at $p_u=1$, the range of $f_c$ exhibits a reduced span 
(about 12\% of the overall population) and the average number of links per
agent stabilizes at $\langle{}l_i\rangle\approx{}13$.
Only an average of 0.5\% of the agents have lost
all their links at a given timestep (and are then randomly rewired).
Decreasing $p_u$ allows for a mean-field approximation of $l_i$, denoted $\langle{}l_i\rangle{}^{mf}$, based
on Eq. (\ref{eqmax}): If  the number of cooperating agents does
not oscillate too strongly, the additional payoff averaged over the neighbours can be roughly estimated 
to be $\langle{}P_i^{add}(t+1)\rangle\approx{}\beta{}Rf_c$. 
Since for the specific form of payoff matrix chosen here
$T-R=I-S=1$  one can
simply add $l_i^c(t)$ and $l_i^d(t)$ in Eq. (\ref{eqmax}) and obtains
$\langle{}l_i\rangle^{mf}\approx{}13$ for $p_u=0.6$ ($f_c\approx{}0.75$).
The actual observed average  of $\langle{}l_i\rangle\approx{}12$ is in 
agreement with this approximation.

We now investigate the dependence of $f_c$ on the chosen update-probability
$p_u$ more closely:
In Figure \ref{bifurc}a we show the dependence of the $f_c$ states visited in
dependence on $p_u$. The plot shows $500$ realizations of $f_c$ (y-axis) 
for different values of $p_u$ (taken after discarding the first $10^3$ steps for
each $p_u$). In Figure \ref{bifurc}a, $p_u$ is thus changed in discrete steps of
$\Delta{}p_u=0.0025$.
One recognizes that the $f_c$-states are not trivially visited
by the system: Although the system is oscillating strongly and periodically
at $p_u=1$ because of deterministic aspects in the evolution, all the points
within the amplitude of the oscillations are visited due to the randomness
introduced at various points (e.g. randomness in the chosen next-to-nearest
neighbours, randomness in the strategy chosen if expected payoffs are equal
for cooperation and defection, etc.). 
Slightly lowering $p_u$
reveals interesting effects on the configuration of the limit cycle (see
the inset of Fig. \ref{bifurc}a): 
One recognizes that the limit cycle first comprises 3 main points between which
the system 'hops' (i.e. it changes from the state with high $f_c$ to the state
with low $f_c$ with one intermediate step and vice versa), then 4 points and
then again 3 points. For some values of $p_u\epsilon[0.965,0.995]$ certain states
between $f_c^{max}$ and $f_c^{min}$ are
never reached. Between $p_u=0.7$ and $p_u=0.6$ the most frequently visited states change from
$f_c^{max}$ and $f_c^{min}$ to the average value of $f_c$ (the limit cycle
vanishes). 
This is also evident from  plotting $f_c(t+1)$
against $f_c(t)$ in Figure \ref{scatter_plots}.
One recognices that decreasing $p_u$ to 0.7 leads to a smaller gap in the attractor; at
$p_u=0.6$ the gap has vanished (not shown). Further decrease in $p_u$ narrows the
space filled by the trajectory of the system (see Figure \ref{scatter_plots},
$p_u=0.1$). 
We have also determined the average of the double amplitude $\langle{}f_c^{max}-f_c^{min}\rangle{}$ of the oscillations,
see Figure \ref{bifurc}b. For each value of $p_u$ investigated we simulated various
realizations of time-series of length $T=10^3$ (discarding the first $10^3$
steps) and averaged the obtained ranges over these
realizations. As Figure \ref{bifurc}b  shows, the range of oscillations reduces for lower
$\alpha=\beta$, as expected. As Figure \ref{bifurc}b shows, we did not find that the amplitude follows a
simple scaling function along the bifurcation, thus suggesting that the
observed change in dynamics is more complicated than a simple Hopf-bifurcation.

As far as the overall dependence of the mean of $f_c$ on $p_u$ is concerned, Figure
\ref{extra_correlation}a shows $\langle{}f_c\rangle{}$ for different values of $p_u$. 
For $\alpha=\beta=6$, the curve exhibits a maximum at $p_u\approx{}0.6$, which 
can be intuitively understood as a trade-off effect between two aspects: On the one
hand decreasing synchronization improves stability and efficiency in the
system as the estimates for the future actions of neighbours become better and
overreaction (extreme oscillations) is reduced. On the other hand, decreasing $p_u$ reduces
efficiency via reducing the reaction of agents to the changes of their neighbours' strategies.
Turning towards the aspect of decreasing $\alpha$ and $\beta$, 
the system is stabilized at a higher value of $p_u$, as the gray line corresponding to $\alpha=\beta=4$ has its
maximum at $p_u\approx{}0.7$. This can be intuitively
understood: As link-dynamics are slowed down for  $\alpha=\beta=4$ 
agents have to be able to react slightly faster to employ efficient internal sanctions via cancellation of links.
Another intuitive reason lies in the fact that lowering $\alpha=\beta$ effectively reduces 
the amplitude of oscillations. One would therefore anticipate 
that the limit cycle vanishes for a higher value of
$p_u$.
We also note that these macroscopic dynamics results in a nice aspect: If agents are
too eager (too fast, high update probability $p_u$) to optimize their
neighbourhood, the global
level of cooperation becomes suboptimal when compared to 
slow adaption ('sloppyness', low $p_u$): If  agents optimize their
situation too 'fast' everybody is worse off on average.

\begin{center}
{\renewcommand{\baselinestretch}{1.0}\normalsize 
\begin{figure}
\begin{center}
\resizebox{0.3\textwidth}{!}{\includegraphics{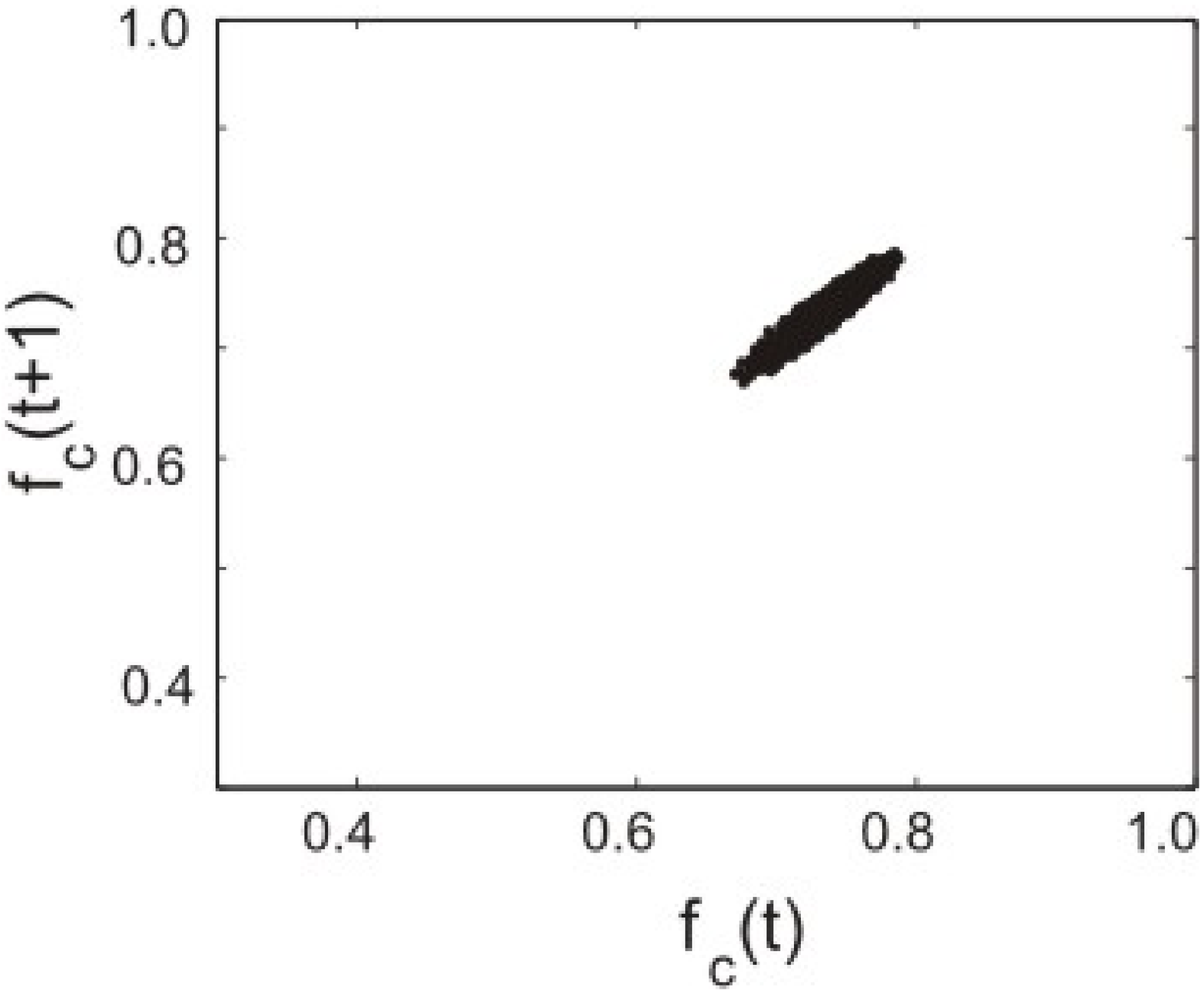}}
\resizebox{0.3\textwidth}{!}{\includegraphics{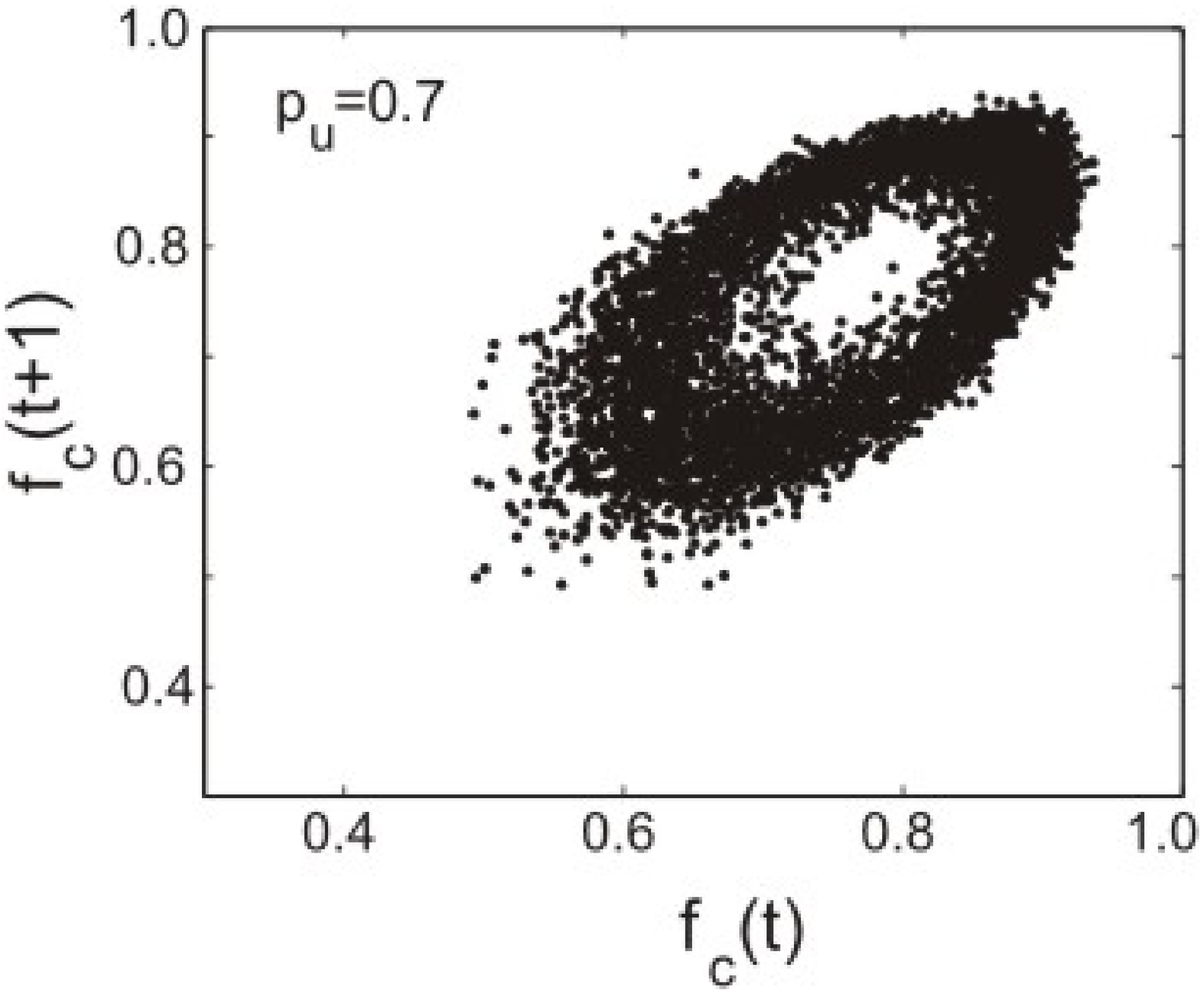}}
\resizebox{0.3\textwidth}{!}{\includegraphics{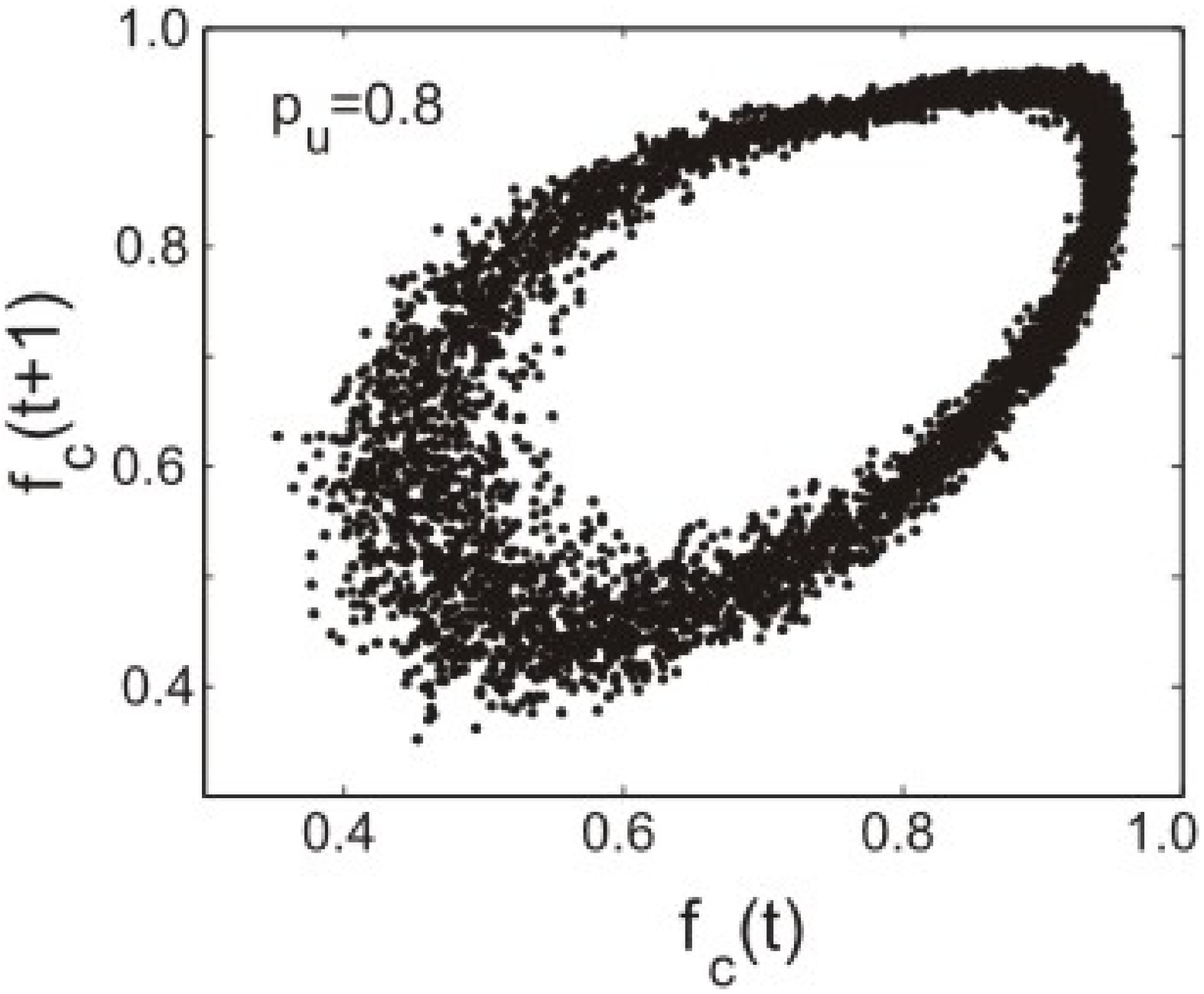}}
\end{center}
\caption{\label{scatter_plots} Visualization of 'attractors' in the space
$\{f_c(t),f_c(t+1)\}$ for different
values of $p_u$ ($p_u=0.1$, $p_u=0.7$, $p_u=0.8$) and $\alpha=\beta=6$.
}
\vspace{1.0cm}
\end{figure}
}
\end{center}

\begin{center}
{\renewcommand{\baselinestretch}{1.0}\normalsize 
\begin{figure}
\begin{center}
\resizebox{1.0\textwidth}{!}{\includegraphics{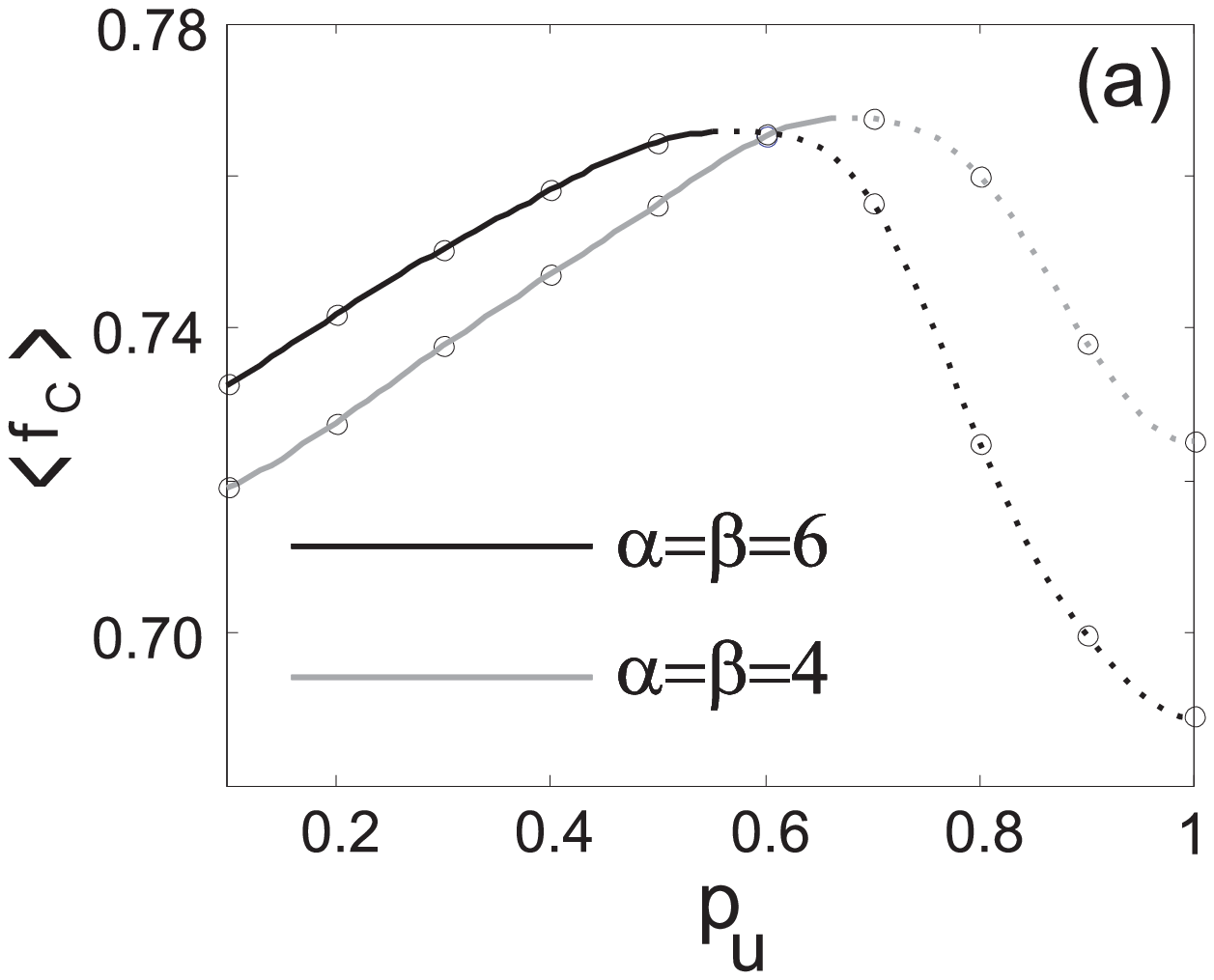}\hspace{1.0cm}\includegraphics{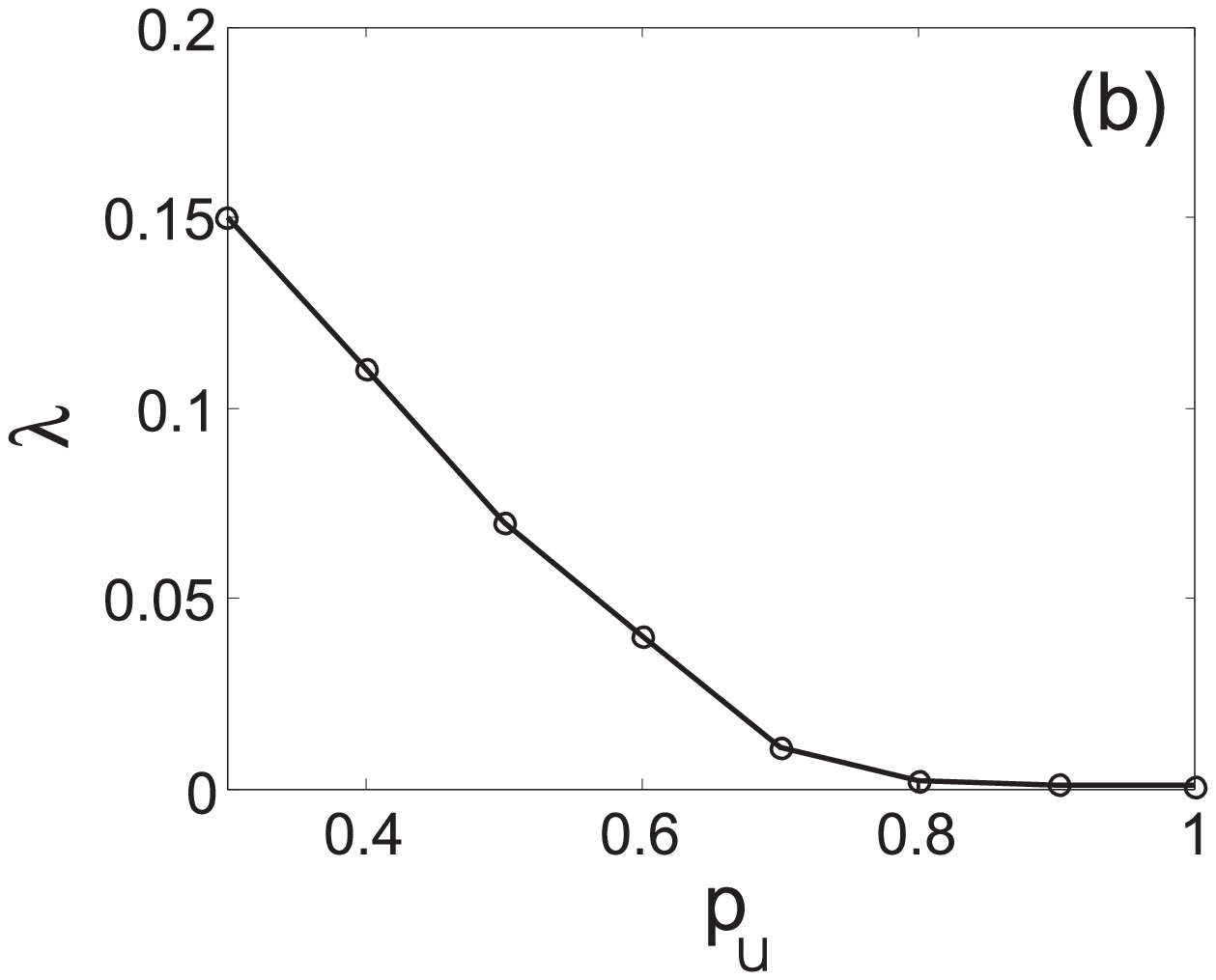}}
\end{center}
\caption{\label{extra_correlation} 
(a) Average number of cooperating
 agents $\langle{}f_c\rangle$ as a
function of update-probability $p_u$ for $\alpha=\beta=6$ and
$\alpha=\beta=4$. (Taking averages of highly correlated time-series is to a certain
extent problematic, which is why the line is drawn broken in the corresponding
regime. To guide the eye, actual values (circles) have been interpolated by a
cubic spline.)
(b) correlation length $\lambda$ 
determined by an exponential fit to the auto-correlation function of
$\Delta{}f_c(t)$ as given in Eq. (\ref{exponential_fit}). 
For very small $p_u$ the exponential fit becomes problematic because the process
becomes practically uncorrelated, i.e. the correlation function turns into
 a Dirac delta
function (the correlation length is not shown for $p_u<0.3$).
}
\vspace{1.0cm}
\end{figure}
}
\end{center}

We have further taken a closer look at the correlations in the system via
the auto-correlation function of the first
differences of $f_c$(t), given by
$\Delta{}f_c(t)\equiv{}f_c(t)-f_c(t-1)$.
The envelope of the auto-correlation function is fitted to an
exponential with inverse correlation length $\lambda$, i.e.
\begin{equation}
\label{exponential_fit}
\langle{}\Delta{}f_c(t+\tau)\Delta{}f_c(t)\rangle\sim{}e^{-\lambda\tau}
\end{equation}
for $\tau>0$. 
Values of $\lambda$ for different
update-probabilities are summarized in Figure
\ref{extra_correlation}b. 
As expected, between $p_u=1$ and $p_u=0.8$
correlation is very strong.
Lowering $p_u$ below 0.8 leads to an
decrease in the correlation, where the exponential fit becomes more and more
problematic. We found that for $p_u<0.3$, the correlation function resembles the shape of a
Dirac delta function and the exponential fit loses sensibility.

Let us now discuss the important point 
of how sensitive the results are to
changes in the specific dynamics chosen.
Towards this end, we have compared results of the model in the
form presented here with a formulation without the $\Theta(p_j(t))$ term in
Eq. (\ref{est_add_payoff}). This variation only leads to slight changes in the
oscillatory states of the system at high $p_u$ (giving a slightly lower 
$\langle{}f_c\rangle$). For lower $p_u$ the difference between the two
formulations became negligible (not shown).
A more massive change in the dynamics occurs when reformulating the model via dropping
the specific assumption of locality, i.e. the assumption of building up new links only to next-to-nearest
neighbours. To do so, we implemented a variant in which 
$\beta$ new links to a set of random nodes
$\tilde{N}^{rand}(t+1;\beta)$ in the system are established. 
The agents now know the
strategies of random players in the system and the payoff of additional links 
is determined by $\bar{P}_i^{add,V1}(t+1)=\sum_{\tilde{N}^{rand}(t+1;\beta)}a_i^cP_{ij}a_j(t)$.
Figure \ref{pic2}c shows the respective dynamics of $f_c(t)$: For
$p_u^{rand}=1$ we recover oscillatory behaviour. Compared to the 
$p_u=1$ case of the initial model we observe a considerably reduced amplitude.
This is expected since $\tilde{N}^{rand}_i(t+1;\beta)$ will always contain
defectors, whereas the initial $\tilde{N}_i(t+1)$ mainly consists of
cooperators (as the next-to-nearest neighbours are typically cooperators since 
links to defectors are immediately cancelled). This can be compared to the
expected payoff being reduced via lowering $p_u$ in the initial model, which
results in more defecting next-to-nearest neighbours 
(as sanctions are not applied immediately).
Apparently, Figure \ref{pic2}c suggests that these two effects are nearly
analogous when choosing the update-probabilities appropriately. Apart from
the observation that the average level of cooperation in the random variant 
is a little bit higher than in the initial model the time-series match 
quite well (see the empirical distribution in the inset). Concerning the
slightly higher mean of the random variant we conjecture that this is due to
a welfare effect stemming from the elimination of imperfect information and
from knowledge about global topology. 
Although the 'random variant' of the model provides a closer
understanding of the model-dynamics, we will continue with the discussion of the initial model since
it is much more realistic.

\subsection{Impact of 'agility' $\alpha$ and $\beta$ and the influence of 
  payoff matrix elements}
 To quantitatively describe the influence of the parameters $\alpha$ and
$\beta$ on $f_c$, we kept both parameters equal and performed
simulations for values ranging from $\alpha=\beta=1$ to $\alpha=\beta=7$. Results
are summarized in
Figure \ref{pic4}a, for $p_u=0.5$ and for the payoff matrix given in Eq. (\ref{payoff_matrix}). Only for
a parametrization of $\alpha=\beta=1$, the majority of agents is defecting. 
One recovers very unstable networks of low average connectivity
of approximately 1.3 links per agent in this case.
For higher values of $\alpha=\beta$, the
system gets initially stabilized and the increase of $f_c$ flattens as
$\alpha$ increases. 
This can be understood as the parameter $\alpha$ has reached a value,
where cooperating agents are able to cancel virtually all the links they have with
defecting agents. In other words, 
the internal sanction-potential of the system has reached a maximum. 
\begin{center}
{\renewcommand{\baselinestretch}{1.0}\normalsize 
\begin{figure}
\begin{center}
\resizebox{1.0\textwidth}{!}{\includegraphics{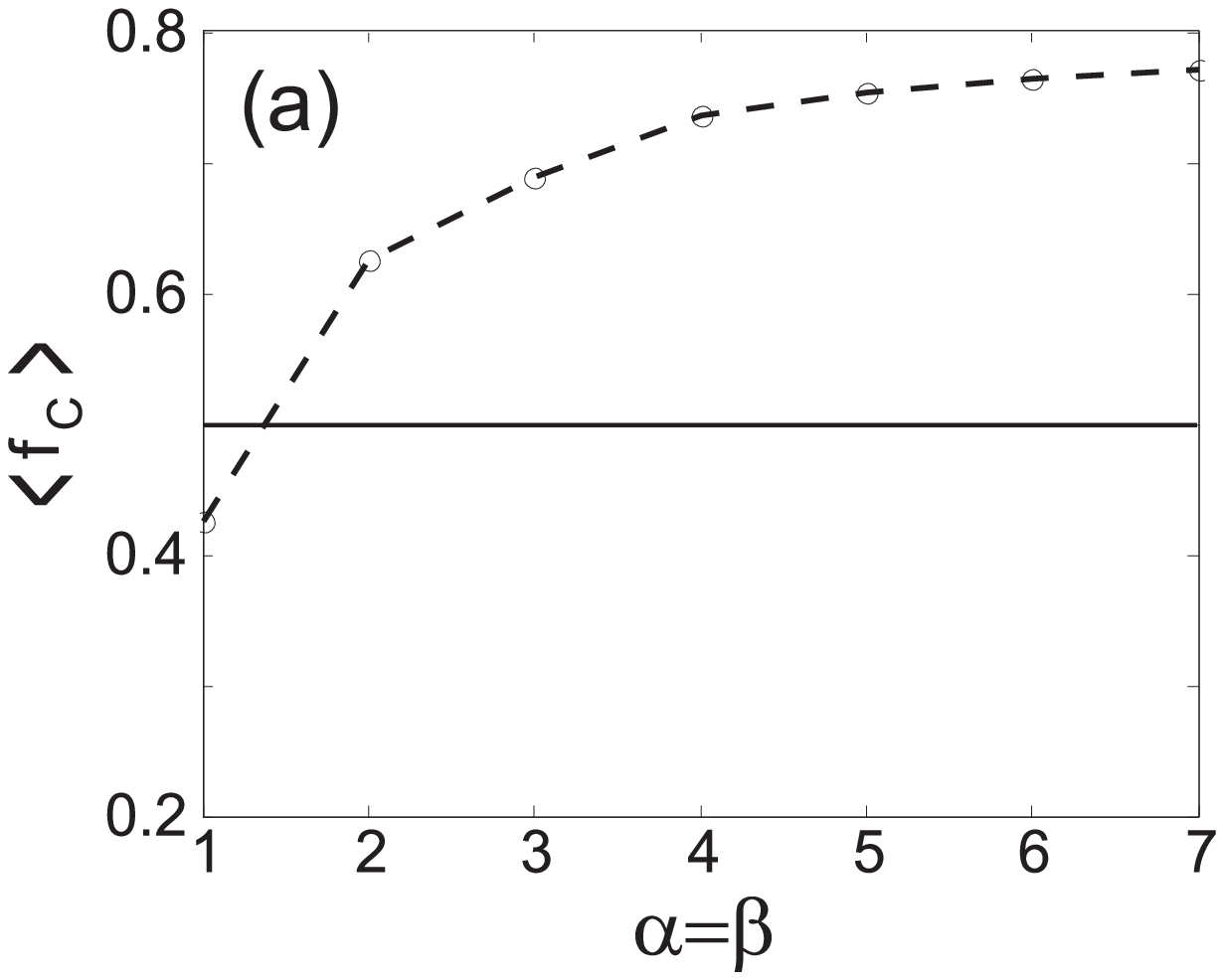}\hspace{1.0cm}\includegraphics{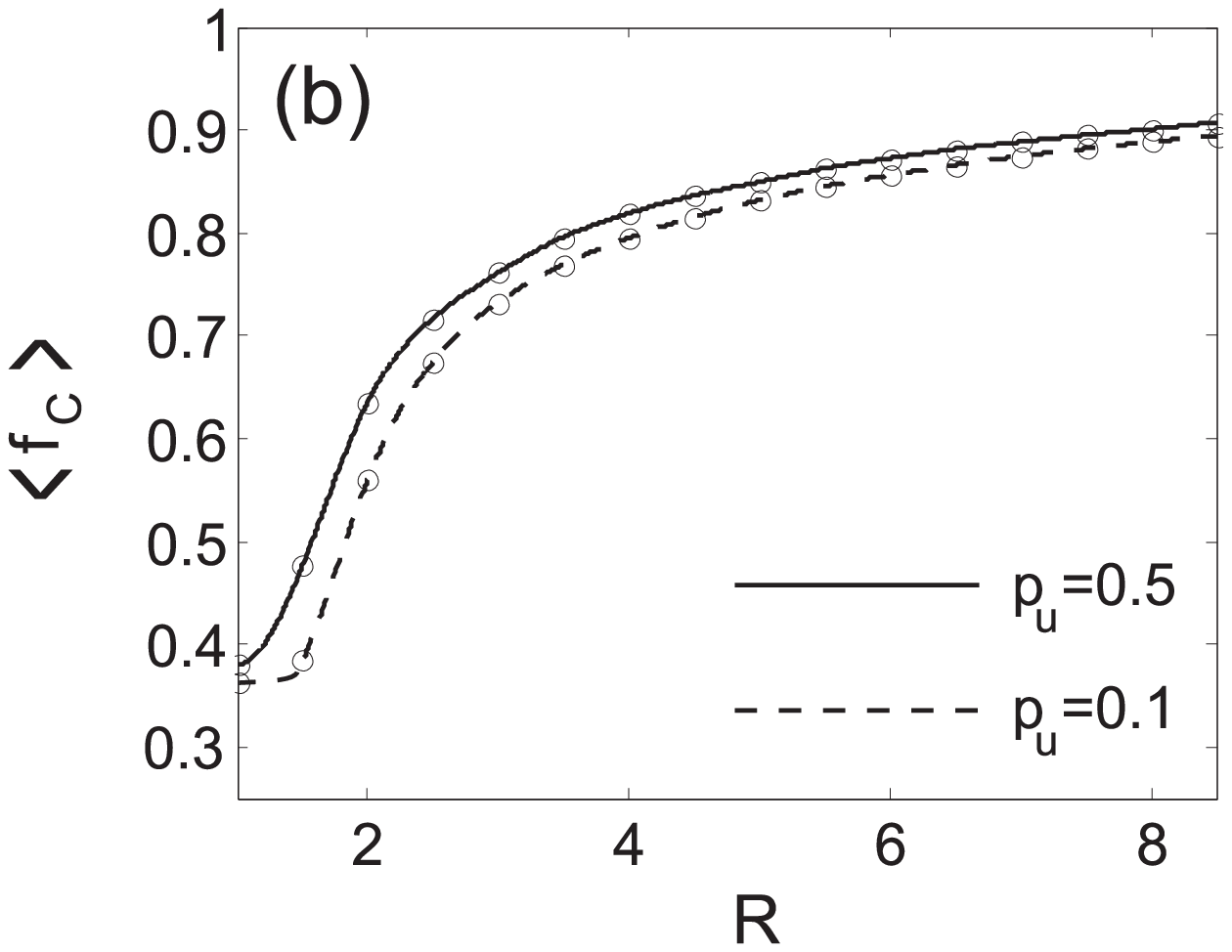}}
\end{center}
\caption{\label{pic4} (a) Influence of parameters $\alpha$ and $\beta$ on the
ratio of cooperating agents in a population of $10^3$ players. The two
parameters have been kept equal. Simulations were done for
$p_u=0.5$.
(b)  Influence of the payoff matrix element $R$ on $\langle{}f_c\rangle$,
where $T=1+R$ and the other elements remain unchanged.
}
\vspace{0.5cm}
\end{figure}
}
\end{center}

 Clearly, not only the parameters $\alpha$, $\beta$ and $p_u$, but also the
 entries in the payoff matrix
$P_{ij}$ influence the dynamics obtained within the presented model.  
In this context, the entry for constellation $I$ (equal to 0 in Eq. (\ref{payoff_matrix}))
is of fundamental importance: When chosen such that 
defecting agents keep the links between one and another, 
a collapse of cooperation in the system is observed.
 On the other hand, increasing the values for temptation $T$ and
reward $R$, while holding their difference $T-R$ constant, 
increases the average number of cooperating agents. This is expected as 
the relative advantage of defectors over cooperators is reduced. 
The corresponding interrelation is quantitatively captured in Figure \ref{pic4}b for 
two values of $p_u$ ($p_u=0.5$ and $p_u=0.1$),
$\alpha=\beta=6$ and $T=1+R$. Again, we observe a saturation effect similar to
the one found when varying $\alpha$ and $\beta$.
Additionally, for 'low' values of $R$, the update-probability apparently has comparatively larger 
influence on $\langle{}f_c\rangle$, whereas in the region of saturation, 
the increase caused by 'switching' between $p_u=0.1$ and
$p_u=0.5$ is decreasing more and more. 

\begin{center}
{\renewcommand{\baselinestretch}{1.0}\normalsize 
\begin{figure}
\begin{center}
\resizebox{0.9\textwidth}{!}{\includegraphics{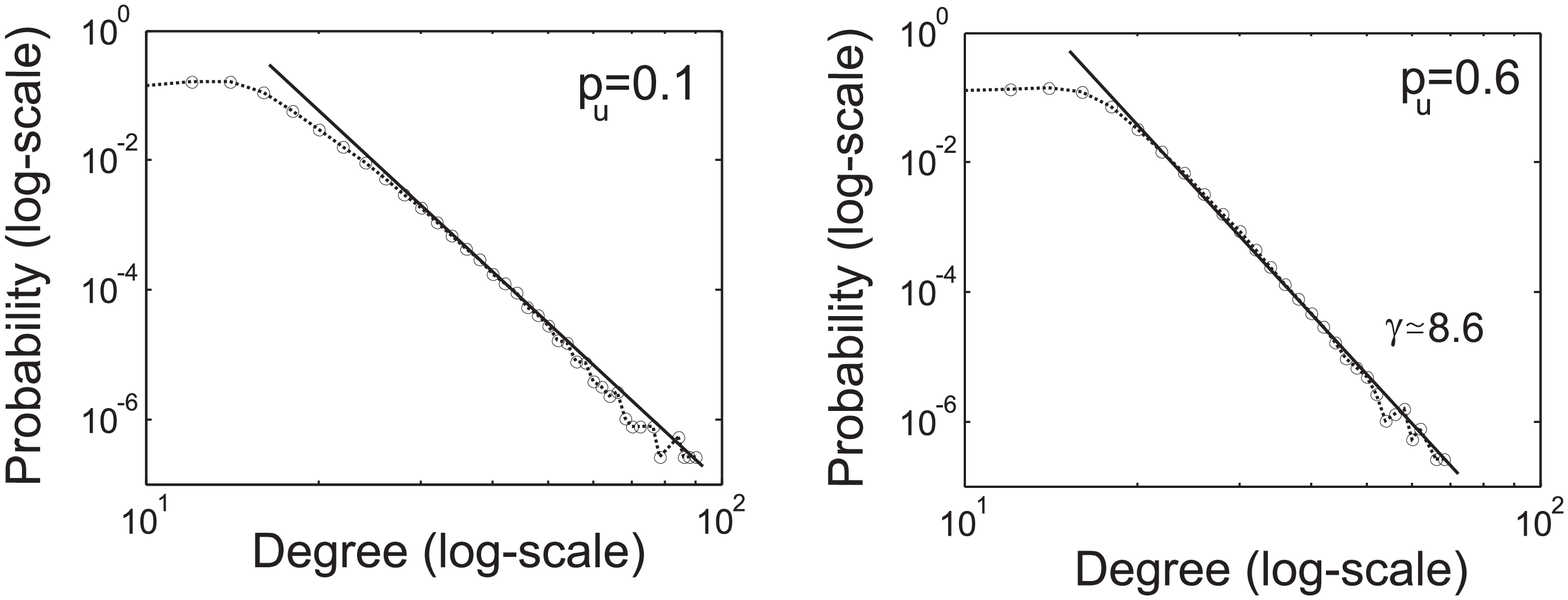}}
\end{center}
\caption{\label{pic5}
Degree distributions averaged over time series with $T=10^5$, $N=10^3$,
$\alpha=\beta=6$ for two different values of the update
probability $p_u$ (0.1 and 0.6). The tail for $p_u=0.6$ may be fitted by a
power-law $P(k)\sim{}k^{-\gamma}$ with $\gamma\sim{}8.6$.}
\end{figure}
\vspace{1.0cm}
}
\end{center}

\subsection{Emerging network topology}
The networks obtained as snap-shots of the dynamics exhibit  
interesting properties, resembling features of real-world networks.
We confine ourselves to the discussion of the two most widely used quantities in the analysis of  
networks, the degree distribution and the cluster-coefficient \cite 
{barabasi,doro}.
Figure \ref{pic5} shows the degree distribution $P(k)$ in a double-logarithmic
 plot for two values of $p_u$. To improve the accuracy of the plot, degree
distributions of networks at $10^3$ different times have been averaged. The
correlation in the time-series has been taken care of by using time-intervals of inverse correlation length.
Figure \ref{pic5} depicts the degree-distribution for the $N=10^3$ case,
for $p_u=0.1$ and $p_u=0.6$. For $p_u=0.1$, the power-law fit shown is slightly
inadequate and indicates a function somewhere between
a power-law and an exponential regime.
For $p_u=0.6$, the double-logarithmic plot
indicates that the tail of the distribution can be expressed as a  
scaling law
$P(k)\sim{}k^{-\gamma}$ with $\gamma\sim{}8.6$.
This shows that the network is clearly not random, but possesses self- 
similar
structure. 
Lowering $p_u$, the network
loses structure and becomes more random. We repeated the analysis for
larger networks ($N=5000$) and did not obtain different results.

\begin{center}
{\renewcommand{\baselinestretch}{1.0}\normalsize
\begin{figure}
\begin{center}
\resizebox{0.45\textwidth}{!}{\includegraphics{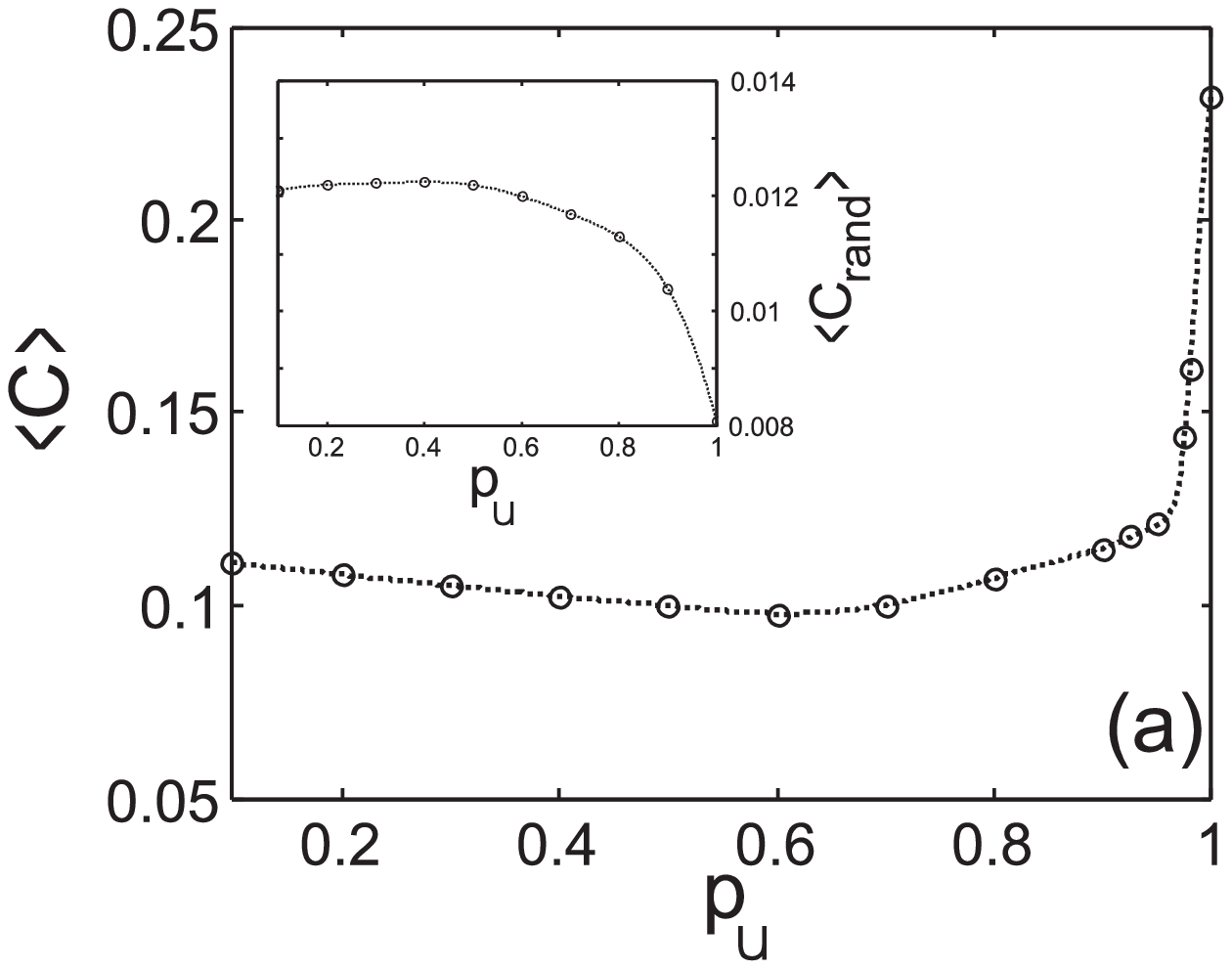}}
\resizebox{0.45\textwidth}{!}{\includegraphics{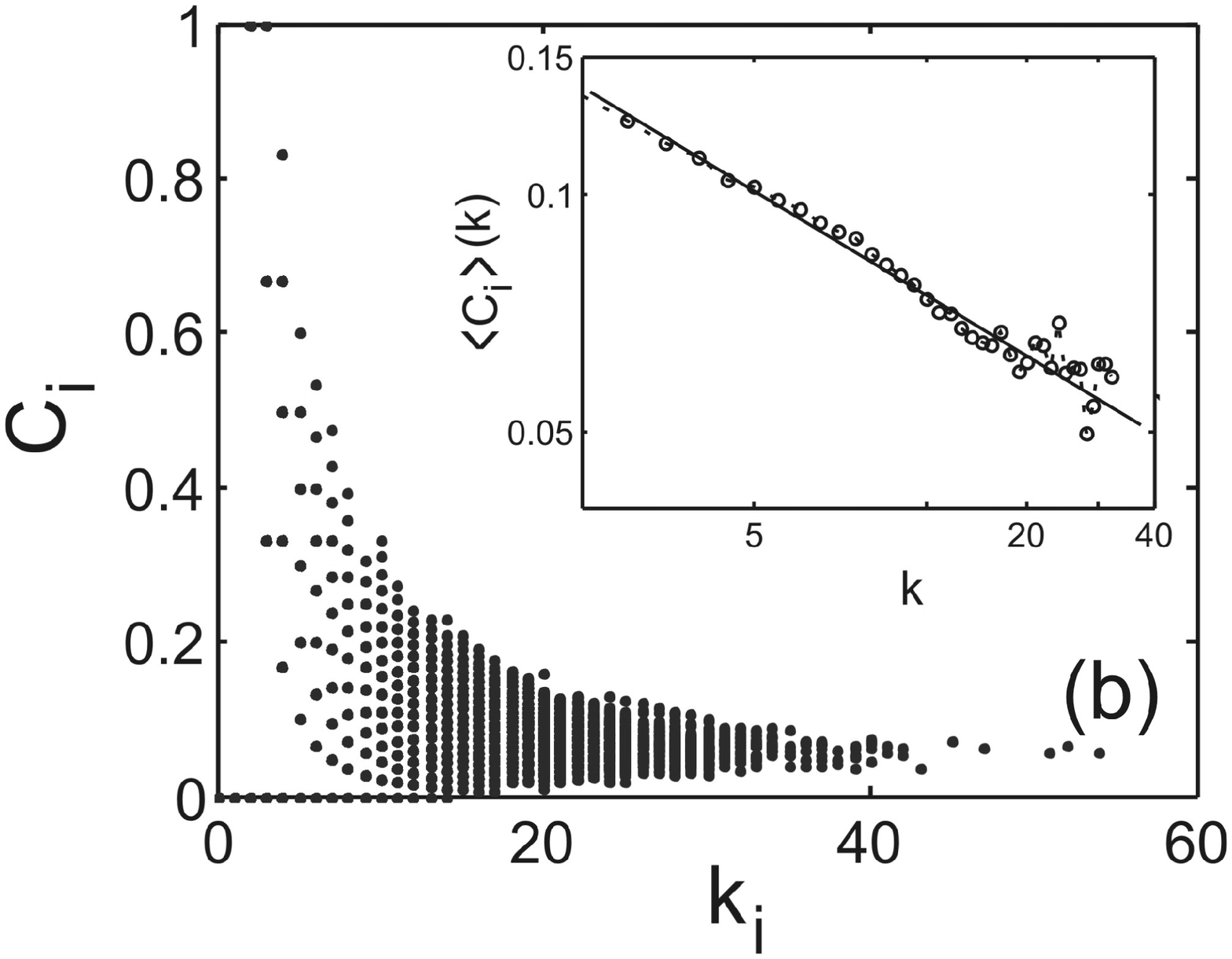}}
\end{center}
\caption{\label{cluster_pic}  (a) Time averages $\langle{}C\rangle$  
of the cluster coefficient for different values
of $p_u$ and $\alpha=\beta=6$. The insert shows the cluster  
coefficients of
equivalent random graphs, denoted by $\langle{}C_{rand}\rangle$.
$\langle{}C_{rand}\rangle$ is decreasing
strongly for $p_u>0.7$ because the average number of links in the  
system is
dropping considerably  in this regime.
(b) Individual cluster coefficients $C_i$ plotted
against individual degree $k_i$ for $p_u=0.6$ and $\alpha=\beta=6$.  
The insert
shows the tail of the corresponding distribution in a double- 
logarithmic plot,
where the individual $C_i$'s
have been averaged. The slope of the interpolating line is $\delta 
\sim{}-0.4$.
}
\end{figure}
}
\end{center}
The cluster coefficient $C$, defined as the average of all individual
cluster coefficients $C_i$, provides a quantitative measure for  
cliques (i.e. circles of acquaintances in the network in which every  
member knows every other
member) in the network. The individual cluster coefficient of a node  
$i$ is defined as
\begin{equation}
C_i=\frac{2E_i}{k_i(k_i-1)}\quad,
\end{equation}
where $E_i$ are the number of existing edges between $i$'s neighbours  
and
$k_i(k_i-1)/2$ gives the highest possible value of edges between the  
neighbours.
As the expected total number of edges in a
random graph can be obtained via  $\bar{l}_{tot}=p(N(N-1))/2$, one  
can compare the
cluster coefficient obtained from given networks to those of  
equivalent random networks.
Figure \ref{cluster_pic}a shows the average cluster coefficients from
simulations at different values of $p_u$, the other parameters being  
kept fixed ($\alpha=\beta=6$).
For comparison, also the cluster coefficients of equivalent random  
graphs are shown
($C_{rand}=p=\langle{}k\rangle/N$).
Obviously, the observed networks exhibit large clustering- 
coefficients when compared to
those of equivalent random graphs. This is not surprising, as our  
mechanism of linkage
directly favours the formation of cliques.
When taking a look at the dependence of the cluster-coefficient on  
$p_u$, a minimum at $p_u=0.6$ can be identified.
Interestingly, this minimum corresponds to the maximum of the number  
of cooperating
agents in Figure \ref{extra_correlation}a,
and to the value of $p_u$ where the degree distribution was best  
fitted with a power-law.

Plotting the cluster-coefficients $C_i$ of individual agents vs. their
degree $k_i$ allows for a more sophisticated analysis of network  
structure. The corresponding plot is shown in Figure \ref{cluster_pic}b, where each  
point corresponds to a pair $\{k_i,C_i\}$. The points have been sampled from 100 different  
networks.   Based on this data, we have calculated the mean cluster-coefficient in
dependence of the degree of the nodes, denoted by $\langle{}C_i\rangle{}(k)$. The tail of this
 distribution is shown in double logarithmic scale in the inset of  
Figure   \ref{cluster_pic}b. Clearly, there is a non-random relationship  
between cluster-coefficient and degree. The underlying  
networks exhibit (complex) hierarchical organisation: For small degrees the
  mean clustering is much higher than for large degrees. We also  
evaluated the number of cooperating agents as a function of degree, finding that
$f_c(k)$ grows with degree (not shown). This confirms our expectation  
that the cooperators are the ones who build up new links and at the same  
time do not suffer from loosing ties.

Finally, Figure \ref{payoff_pic} shows the average distribution of
individual payoffs in the system for $p_u=0.1$ and $p_u=0.6$.
Although the maximum of the distribution is at a higher payoff
for $p_u=0.1$, the average payoff is higher for $p_u=0.6$ as the tail of the
distribution is 'fatter' in this case. The inset shows the tails of
the distribution in a semi-logarithmic plot, indicating that the tails are
close to exponential.

\subsection{Experimental evidence for cooperation networks}

As a proxy for a cooperation network of humans it is reasonable to  
consider telephone call networks. It is reasonable to assume that  
communication between cooperating individuals will dominate the total number of calls,  
while non-cooperating individuals will avoid communication. There exists recent research  
on real mobilephone-call networks  \cite{mobile}. In this study, a power-law degree distribution  with a characteristic
exponent $\gamma^{mobile}\sim{}8.4$ was found. It is obvious, that the exponent obtained within our model ($\gamma 
\sim{}8.6$) shows  close resemblance to this value.
This suggests that our model captures dynamics of real-world networks  
and has some predictive value. 
We think that the experimental procedure (temporally  clearly  
limited measurements of networks) behind the data reported in \cite{mobile} is much more comparable  
to the averaging procedure  in our simulations than the procedures behind many other  
investigations hitherto, which often involve effects of growth.

\section{Summary}

In this work, we have considered the prisoner's dilemma being played on
dynamic networks under the assumptions of rationality and strictly local information horizons of the
agents. The novelty lies in the fact that links in the network are treated as
a dynamical variable while -- at the same time -- we  adopted an update-scheme based on
profit-maximization and not on imitation. 
The network on which the game is played is thus an emergent 
structure, co-evolving with the configuration of strategy-space.
Within this framework,
 reasonable assumptions about fully rational individual decisions lead to a model of network dynamics 
where defectors are effectively sanctionized in two ways: By implicitly being affected by link-cancellations and
by explicitly not being able to establish new links as the players minimize
potential losses by accepting only 'recommended' co-players.

We showed that the dynamics implied are non-linear and 
lead to the emergence of cooperative behaviour even within a framework of
rationality.
More precisely, we observed distinct modes in the model: In the case of
high synchronization  of the agents' decisions, significant oscillations of global parameters appear and much
resources are wasted in collective movements. We have discussed the
dependence of the system on the control-parameter in this regime.
For low synchronisation of the agents, randomness in the system and delay of the players reactions reduces cooperation.
For regimes in between high and low synchronization, we showed that the system reaches an optimum, 
where network characteristics  resemble those of complex networks, exhibiting 
 clearly non-random properties like power-law degree distribution and
 hierarchical clustering. Towards this end it is especially remarkable that
 our model predicts a rather high tail-exponent $\gamma=8.6$ of real
 world communication networks (compare with $\gamma^{mobile}=8.4$ in \cite{mobile}).
\begin{center}
{\renewcommand{\baselinestretch}{1.0}\normalsize
\begin{figure}
\begin{center}
\resizebox{0.5\textwidth}{!}{\includegraphics{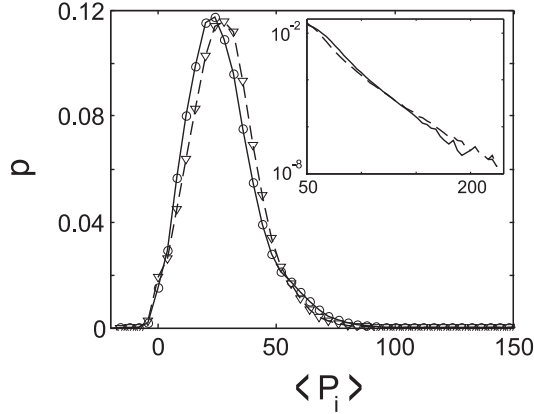}}
\end{center}
\caption{\label{payoff_pic} Average distribution of the individual payoffs
for $p_u=0.1$ (broken line) and $p_u=0.6$ (solid line) ($\alpha=\beta=6$). The
insert shows the tails of the distribution in a semi-logarithmic plot.}
\end{figure}
}
\end{center}

It is interesting that oscillatory dynamics immanent for high synchronization
have also been found in a spatial formulation of the prisoner's dilemma where
participation in the game was voluntary \cite{szabo1}. Thus it 
seems that the cyclical dominance of the strategies found in
\cite{szabo1} can be qualitatively confirmed even within the picture of
(highly) dynamic networks. 
The fraction of cooperating agents in our model was found to be bound by rougly
$f_c^{+}\approx{}0.9$ from above and by roughly $f_c^{-}\approx{}0.4$ from
below, showing a saturation regarding the studied parameters towards $f_c^{+}=0.9$.
This is above the level of cooperation found in the voluntary formulation of the PD
\cite{szabo1} and in the initial work of Nowak \& May \cite{NowakandMay}, 
but below typical fractions found for the PD on variable
networks with imitative behavior of agents \cite{zimmermann04}. 
It is not surprising that imitation on dynamic networks yields higher overall
degree of cooperation than rationality since on fixed structures cooperation
is sustainable for imitation but not (or much less) sustainable for rational
settings.

The current work may be extented in various
directions: On the one hand, we expect that introduction of  
heterogeneity in the payoff matrix and the parameters
$\alpha$ and $\beta$ (i.e. that these parameters take different  
values for the
agents) could lead to further interesting results. We also
conjecture that coupling $\beta$ to some measure of payoff
(fitness) of the individual agents should introduce some new, realistic
effects.

{\bf Acknowledgements}\newline{}
\newline{}
The authors acknowledge funding through the Austrian Science Fund (FWF),
P17621-G05 and P19132-N16. S.T. would like to thank the SFI and in particular J.D. Farmer for the great
hospitality.


\end{document}